\documentclass[cernpreprint,texlive=2011,txfonts,UKenglish]{latex/atlasdoc}
\pdfoutput=1
\usepackage[subfigure=true,biblatex=false]{latex/atlaspackage}
\usepackage{latex/atlasphysics}

\usepackage{multirow}
\usepackage{rotating}
\usepackage{xtab}
\usepackage{cite}
\usepackage{siunitx}
\DeclareSIUnit\fbsiunit{fb}
\usepackage{dcolumn}

\usepackage{listings}
\usepackage{chngcntr}

\usepackage{COMMON-combined}
\newcommand{\intluminoerr}       {20.3\invfb}

\renewcommand*{\TeV}{\ifmmode {\mathrm{\ Te\kern -0.1em V}}\else
	                   \textrm{Te\kern -0.1em V}\fi}%
\renewcommand*{\GeV}{\ifmmode {\mathrm{\ Ge\kern -0.1em V}}\else
	                   \textrm{Ge\kern -0.1em V}\fi}%

\graphicspath{{logos/}{figures/}}
\newcommand{\papertitle}{
Measurement of \Wboson boson angular distributions in events with high transverse momentum jets at \sqseight using the ATLAS detector
}
\AtlasTitle{\papertitle}
\AtlasAbstract{The \Wboson boson angular distribution in events with high transverse momentum jets is measured using data collected by the ATLAS experiment from proton--proton collisions at a centre-of-mass energy \sqseight at the Large Hadron Collider, corresponding to an integrated luminosity of \intluminoerr. The focus is on the contributions to \Wjets processes from real \Wboson emission, which is achieved by studying events where a muon is observed close to a high transverse momentum jet. At small angular separations, these contributions are expected to be large. Various theoretical models of this process are compared to the data in terms of the absolute cross-section and the angular distributions of the muon from the leptonic \Wboson decay.}

\AtlasRefCode{STDM-2015-16}
\AtlasJournal{Phys. Lett. B}
\AtlasJournalRef{Phys. Lett. B 765 (2017) 132}
\AtlasDOI{10.1016/j.physletb.2016.12.005}
\PreprintIdNumber{CERN-EP-2016-182}

\hypersetup{pdftitle={CERN-EP-2016-182},pdfauthor={The ATLAS Collaboration},pageanchor=false}
\author{The ATLAS Collaboration}

\begin{document}

\maketitle

\section{Introduction}
\label{sec:introduction}

Precision measurements of Standard Model processes at the Large Hadron Collider (LHC) are crucial for probing the fundamental structure of the strong and electroweak interactions. The data sample corresponding to an integrated luminosity of \intluminoerr collected by the ATLAS experiment from proton--proton ($pp$) collisions at a centre-of-mass energy \sqseight at the LHC allows detailed study of perturbative quantum chromodynamics (perturbative QCD, pQCD) and real and virtual electroweak (EW) corrections that impact measurements of \Wjets production.

At high energies, real emission of weak bosons in dijet events can contribute significantly to inclusive \Wjets measurements~\cite{Baur:2006sn,Christiansen:2014kba,Krauss:2014yaa,Boughezal:2015oga,Boughezal:2016dtm}. In leading-order (LO) calculations of $\Wboson +  1$-jet production, the \Wboson boson is balanced by the recoil hadronic jet, often referred to as \textit{back-to-back} production. At next-to-leading order (NLO), QCD and EW corrections to $\Wboson +  1$-jet processes appear, both as real and virtual contributions. In the case of real \Wboson boson emission from an initial- or final-state quark, these contributions scale as $\order{\alpha \ln^2 p_{{\rm T},j}/m_\Wboson}$, where $\alpha$ is the gauge coupling of the unified EW theory, $p_{{\rm T}, j}$ is the transverse momentum of the jet and $m_\Wboson$ is the \Wboson boson mass, and have a collinear enhancement in the distribution of the angular distance between the \Wboson boson and the closest jet. The collinear enhancement arises from collinear and infrared divergences which would be present in the limit of vanishing \Wboson boson mass, but which are regulated by its finite mass. The procedures to correctly account for collinear parton radiation, such as massless gluon emission, are well known and led to the introduction of (Sudakov) parton showering of the initial- and final-state partons in Monte Carlo generators for QCD as well as quantum electrodynamics (QED) contributions. An analogous procedure is available for the emission of real \Wboson bosons~\cite{Christiansen:2015jpa}. The effect of real \Wboson boson emission can be probed by isolating events for which the cancellation between real and virtual corrections is incomplete, for example by studying the region of small angular separation between a jet and the \Wboson boson. This region also contains LO contributions from $\Wboson +  2$-jets, as well as corrections to that process, which must be included for accurate predictions.

Due to this complex mixture of $\Wboson +  1$-jet and $\Wboson +  2$-jet processes, and the relevant QCD and EW corrections to both, comparisons of measurements to predictions using multiple approaches for estimating those corrections are crucial. Comparisons of the measured angular spectra of the muon from the \Wboson boson with fixed-order predictions at NLO and next-to-next-to-leading-order (NNLO) and with programs with electroweak parton showers help in understanding the accuracy of these predictions.

The measurements presented here focus on events that contain a muon and a jet with transverse momentum $\pt > 500\GeV$.
In this kinematic regime, contributions to \Wjets processes from real \Wboson boson emission are enhanced in the region of small angular separation between the \Wboson boson decay products and the closest jet.
The angular separation is defined as the distance between the muon and the closest jet, $\DeltaR (\mu, {\rm jet}) = \sqrt{(\Delta \phi)^2 + (\Delta \eta)^2}$,\footnote{ATLAS uses a right-handed coordinate system with its origin at the nominal interaction point (IP) in the centre of the detector and the $z$-axis along the beam pipe. The $x$-axis points from the IP to the centre of the LHC ring, and the $y$-axis points upward. Cylindrical coordinates $(r,\phi)$ are used in the transverse plane, $\phi$ being the azimuthal angle around the $z$-axis. The pseudorapidity is defined in terms of the polar angle $\theta$ as $\eta=-\ln\tan(\theta/2)$.} hereafter referred to as $\DeltaR$.
Measurements of this angular separation thus provide precision tests of pQCD and electroweak predictions for the rate and pattern of real \Wboson boson emission.
Real \Wboson boson emission, also termed collinear \Wboson production, is the dominant process for events with $\Delta R < 2.4$, and thus $\DeltaR < 2.4$ is referred to as the collinear region. The significance of this higher-order contribution at small $\DeltaR$ is shown in Ref.~\cite{Boughezal:2016dtm}. For events with $\DeltaR > 2.4$, the \Wboson boson is balanced by a hadronic recoil that may consist of one or more jets.

These measurements of the $\DeltaR$ distribution probe a new region of phase space that has not been explicitly studied in detail. Measurements of \Wjets production by both the ATLAS and CMS experiments often remove portions of the collinear region by requiring that the lepton ($e$ or $\mu$) is separated from any jet by an angular distance of $\DeltaR > 0.5$~\cite{STDM-2012-24,Khachatryan:2014uva}. By relaxing this requirement to $\DeltaR > 0.2$ and focusing on the distribution of angular separation between the muon and the closest jet in events with at least one very high \pT\ jet ($\pT > 500\GeV$), it is possible to explicitly target real \Wboson emission with this measurement.

Collinear \Wboson production may constitute an important background in searches for beyond the Standard Model physics that involve Lorentz-boosted top quarks~\cite{Rehermann:2010vq}, either in rare topologies or at high energies. If the \Wboson decay products are collinear with one of the jets, the structure of that jet can begin to resemble that of the three-pronged structure of a boosted top quark. While the rate for collinear \Wboson production is suppressed relative to dijet production with no \Wboson emission, hadronic \Wboson decays can cause a large increase in the measured jet mass. The result is that \Wboson emission from quarks at very high \pt can yield single jets with definite substructure that resemble the boosted top-quark signals being searched for.

\section{The ATLAS detector}
\label{sec:Detector}
The ATLAS detector~\cite{PERF-2007-01,PERF-2010-01} provides nearly full solid angle coverage around the $pp$ collision point at the LHC.

The inner detector (ID) comprises a silicon pixel tracker closest to the beamline, a microstrip silicon tracker, and a straw-tube transition-radiation tracker at radii up to $108$ cm. A thin solenoid surrounding the tracker provides a $2~\rm T$ axial magnetic field enabling the measurement of charged-particle momenta. The overall ID acceptance spans the full azimuthal range in $\phi$, and the range $|\eta|<2.5$ for particles originating near the nominal LHC interaction region~\cite{STDM-2010-06}.

The electromagnetic (EM) and hadronic calorimeters are composed of multiple subdetectors spanning $|\eta|<4.9$. The EM barrel calorimeter uses a liquid-argon (LAr) active medium, together with lead absorbers, and covers $|\eta| < 1.45$. In the region $|\eta| < 1.7$, the hadronic calorimeter is constructed from steel absorber and scintillator tiles and is separated into barrel ($|\eta|<1.0$) and extended-barrel ($0.8<|\eta|<1.7$) sections. The endcap ($1.375<|\eta|<3.2$) and forward ($3.1<|\eta|<4.9$) regions are instrumented with LAr calorimeters for EM as well as hadronic energy measurements.

A muon spectrometer with three large air-core toroid magnet systems surrounds the calorimeters. The muon spectrometer measures the momentum of muons from their tracks, which are reconstructed with three layers of high-precision tracking chambers. These chambers provide coverage in the range $|\eta|<2.7$, while dedicated fast chambers allow triggering in the region $|\eta|<2.4$.

A three-level trigger system is used to record events for analysis. The different parts of the trigger system are referred to as the Level-1 trigger, the Level-2 trigger, and the Event Filter~\cite{PERF-2011-02}. The Level-1 trigger is implemented in hardware and uses a subset of detector information to reduce the event rate to a design value of at most 75~kHz. The Level-1 trigger is followed by two software-based triggers, the Level-2 trigger and the Event Filter, which together reduce the event rate to a few hundred Hz.

\section{Data and simulated samples}
\label{sec:datamc}
The measurement presented here is based on the entire 2012 \pp dataset at a centre-of-mass energy of \sqseight. Events are required to meet baseline quality criteria during stable LHC running periods. These data quality criteria primarily reject data with significant contamination from detector noise or issues in the read-out~\cite{LARG-2013-01} based upon individual assessments for each subdetector. The resulting dataset corresponds to an integrated luminosity of \intluminoerr.
The absolute luminosity scale is derived from beam-separation scans performed in November 2012.
The uncertainty in the integrated luminosity is $\pm 1.9\%$~\cite{DAPR-2013-01}.

Simulated events from Monte Carlo (MC) generators are used for calculating the signal efficiency and estimating background in the signal region.
The events are simulated using a GEANT4-based~\cite{Geant4} full detector simulation~\cite{SOFT-2010-01-bugfix}.
In addition to the hard scatter, each event is overlaid with a number of additional $pp$ collisions (\pileup) extracted from the distribution of the average number of \pp interactions per bunch crossing $\mu$ observed in data.
These additional $pp$ collisions are generated with \Pythia~v8.160~\cite{Pythia8} using the ATLAS A2 set of tuned parameters (A2 tune)~\cite{MC12AU2} and the MSTW2008LO parton distribution function (PDF) set~\cite{Martin:2009iq}.

Events containing \Wjets are generated with \Alpgen~2.14~\cite{alpgen}, which implements MLM matching~\cite{Mangano:2006rw} of the matrix element calculation with parton showering.
The \Wboson boson is produced as part of the matrix element calculations, allowing simulation of both collinear and back-to-back \Wjets production.
In the latter, the \Wboson boson is balanced by the hadronic recoil system.
The matrix elements provided by \Alpgen are configured to allow up to five partons in the final state in addition to the \Wboson boson, including heavy-flavour production as well.
The generator is interfaced with \Pythia~v6.427~\cite{pythia} for parton showering and fragmentation.
The CTEQ6L1 PDF set~\cite{PDF-CTEQ} is used.
A $K$-factor is applied to these samples to correct the normalisation to a NNLO pQCD inclusive cross-section calculated with FEWZ~\cite{Melnikov:2006kv} and the MSTW2008NNLO PDF set.
A sample of events is also generated with \Pythia~v8.210 and using the CT10 NLO PDF set~\cite{Lai:2010vv} in which \Wboson boson radiation can be produced via a weak parton shower.

Dijet events are generated with \Pythia~v8.165.
Top-quark pair production is simulated with \Powheg-r2129~\cite{Nason:2004rx, Frixione:2007vw, Alioli:2010xd, Frixione:2007nw} interfaced with \Pythia~v6.426 with the P2011C~\cite{Perugia2010} tune for parton showering and fragmentation.
Diboson production is simulated with \MCatNLO~v4.07~\cite{mcatnlo}.
Additional samples of diboson production are generated using \Sherpa~v1.43~\cite{Gleisberg:2008ta} and these are used to estimate theoretical uncertainties in the diboson background estimation.
The above samples are all generated using the CT10 NLO PDF set.
Events containing $Z+$ jets are generated with \Alpgen using the same configuration as the \Wjets simulation above.
Single top-quark production is a negligible background for this analysis and is not included.

All samples are normalised to their calculated inclusive cross-sections.
However, for the \Wjets, dijets, \ttbar and \Zjets samples, there is an additional correction applied to the normalisation, derived from the comparison of data and Monte Carlo simulations in the signal region and control regions.
The process of deriving this correction is explained in detail in \secref{event-selection}.

\section{Object and event selections}
\label{sec:event-selection}
\subsection{Baseline event selection}
\label{sec:event-selection:baseline}

The topology of collinear \Wboson production involves two back-to-back high-\pt jets, one of which emits a nearby \Wboson boson.
Events are required to contain at least one jet with $\pT > 500\GeV$, as this is found to be sufficient to probe the kinematic region of interest.
The probability of a collinear \Wboson emission from such a jet is estimated by \Pythia~v8.210 to be $0.15\%$.
Over half of the production of \Wjets in the phase space probed in this measurement is in the collinear region.
A requirement for a second high-\pt\ jet is not applied.
Although both jets initially recoil from each other and have similar \pt, the jet that emits the collinear \Wboson boson can lose a significant amount of energy to the muon and neutrino, neither of which are reconstructed as part of the jet energy.
Requiring a second high-\pt\ jet would impose an implicit maximum on the energy carried by the \Wboson boson and its decay products.

The analysis focuses on the leptonic decays of \Wboson bosons to muons in order to ensure a high reconstruction purity, and thus events are required to have exactly one muon.
Events that contain an electron are rejected, which reduces the background by removing mixed-flavour dileptonic (electron plus muon) \ttbar decays. Control regions are used to establish the normalisation of MC simulations of several background processes.
These regions are defined by inverting various selection criteria used in the final measurement.

To reject non-collision background~\cite{DAPR-2012-01}, events are required to contain at least one primary vertex consistent with the beam-interaction region, reconstructed from at least two tracks each with $\pttrk>400$~MeV. The primary hard-scatter vertex is defined as the vertex with the highest $\sum(\pttrk)^2$. To reject rare events contaminated by spurious signals in the detector, all \antikt~\cite{Cacciari200657,Cacciari:2008gp} jets with radius parameter $R=0.4$ and $\ptjet>20 \GeV$ (see below) are required to satisfy the loosest jet-quality requirements discussed in Ref.~\cite{DAPR-2012-01}. These criteria are designed to reject non-collision background and significant transient noise in the calorimeters while maintaining an efficiency for good-quality events greater than 99.8\% with as high a rejection of contaminated events as possible. In particular, this selection is very efficient in rejecting events that contain fake jets due to calorimeter noise.

\subsection{Trigger selection}
\label{sec:event-selection:trigger}

Events used in this analysis are selected by requiring that they pass at least one of two single-muon triggers~\cite{TRIG-2012-03}.
The first trigger requires an isolated muon with $\pt > 24\GeV$ and the second trigger requires a muon with $\pt > 36\GeV$ with no isolation criteria applied.
The track-based isolation used in the trigger requires that the scalar sum of the \pt of all tracks within a cone of radius $\Delta R = 0.2$ around the muon is less than $12\%$ of the muon \pt.

\subsection{Object reconstruction}
\label{sec:event-selection:objects}

Muons are reconstructed by combining tracks in the ID with tracks in the muon spectrometer~\cite{PERF-2014-05}.
They are required to have $\pt > 25\GeV$ and $|\eta| < 2.4$.
To reduce contamination from semileptonic $b$-decays, in-flight pion and kaon decays and cosmic muons, their longitudinal impact parameter with respect to the primary vertex $z_0$ must satisfy $|z_0|\sin\theta < 0.5$ mm and their transverse impact parameter with respect to the primary vertex $d_0$ must satisfy $|d_0|/\sigma(d_0) < 3$.
The selected offline reconstructed muon must also match the online muon that passed the trigger.

Jets are built using the \antikt\ algorithm with a radius parameter of $R=0.4$ from locally calibrated three-dimensional topological energy clusters~\cite{PERF-2014-07}.
The resulting jets are required to have $\pt > 100\GeV$ and $|\eta| < 2.1$.

The number of $b$-tagged jets for a given event is calculated using the MV1 tagger~\cite{PERF-2012-04} on jets built using the \antikt\ algorithm with $R=0.4$.
The jets considered for $b$-tagging have $\pt > 25\GeV$ and are reconstructed within $|\eta| < 2.1$.
The MV1 tagger is configured to have a $b$-tagging efficiency of 70\% in semileptonic \ttbar\ events.

Electrons are reconstructed from a combination of a calorimeter energy cluster and a matched ID track~\cite{PERF-2013-03,PERF-2013-05}.
They must meet a set of identification criteria (the so-called \textit{medium} criteria of Ref.~\cite{PERF-2013-03}).
They are also required to have $\pt > 20\GeV$ and $|\eta| < 2.47$, excluding the transition region between the barrel and the endcap calorimeters ($1.37 < |\eta| < 1.52$).
To reduce the contamination from semileptonic $b$-decays and misidentification, the same impact parameter requirements used for muons are applied along with an isolation requirement.
This isolation is track-based and requires that the scalar sum of the \pt of all tracks in a cone of radius $\Delta R = 0.2$ around the electron be less than $15\%$ of the electron \pt.

\subsection{Measurement selection}
\label{sec:event-selection:signal}

To select the \Wjets signal, events are required to contain at least one jet with $\pT > 500\GeV$, exactly one muon, no $b$-tagged jets, a primary vertex and no electrons.
Any additional jets with $\pt > 100\GeV$ are included in the analysis.
The leading jet, defined as the jet with the highest \pt, is not necessarily the one closest to the muon. The $\Delta R$ distance is always measured with respect to the closest jet.
The muon is required to be isolated using both track-based and calorimeter-based isolation criteria.
The track isolation requires that the scalar sum of the \pt of all tracks in a cone of radius $\Delta R = 0.2$ around the muon be less than $10\%$ of the muon \pt.
The calorimeter isolation requires that the scalar sum of the \pt in all calorimeter cells in a cone of radius $\Delta R = 0.2$ around the muon be less than $40\%$ of the muon \pt.
Applying these isolation criteria significantly reduces the background from dijet events, where muons mostly originate from heavy-flavour or in-flight decays and are non-isolated.
The $b$-tag veto also reduces the background from \ttbar, which generates two $b$-quarks in their decay, by over 80\%, while only 10\% of the \Wjets signal is rejected.
Requirements on missing transverse momentum were not found to improve the signal selection or background rejection.
The efficiency of the isolation requirement was studied both in simulated samples and in situ using data events containing high-\pt\ top quarks, and the results from the two studies were in agreement.
However, in the extremely collinear region, where the distance between the muon and the closest jet is $\DeltaR < 0.2$, the limited size of the event sample did not allow the same conclusion.
As a result, events where $\DeltaR < 0.2$ are also excluded.
This causes approximately 2\% of the \Wjets signal to be rejected.

\subsection{Control region definitions and background estimation}
\label{sec:event-selection:background}

For the final state with at least one high-$\pt$ jet and a single muon, the dominant background processes that contribute to the signal region are dijets, $\ttbar$ and \Zjets.
In addition, there is a small background contribution from diboson production.
These are all modelled using the simulated samples described in \secref{datamc}.

For each of the three main background processes, a control region utilising an event selection different from the signal region is defined such that most of the events in this control region are from the chosen background.
Control Region 1 is enriched in dijets, with a 93\% purity of dijet events, by applying the inverse of the signal region isolation selection.
It uses events that pass the muon trigger without an isolation requirement and requires the muon to have $\pt > 38\GeV$, as events with a non-isolated muon of lower \pt are mostly rejected by the trigger, together with a distance $\Delta R > 0.2$ between the muon and the closest jet.
Control Region 2 is enriched in \ttbar, with 91\% of events originating from \ttbar production, by requiring at least two $b$-tagged jets.
Control Region 3 is enriched in \Zjets, which constitute 94\% of events in this region, by using events with exactly two muons, with both muons passing the signal region isolation.
It is further required that the dimuon invariant mass in Control Region 3 satisfies $60\GeV < m_{\mu\mu} < 120\GeV$.
In this case, the muon with the higher \pt is chosen to define $\DeltaR$.

\begin{figure}
  \centering
  \includegraphics[width=3.1truein]{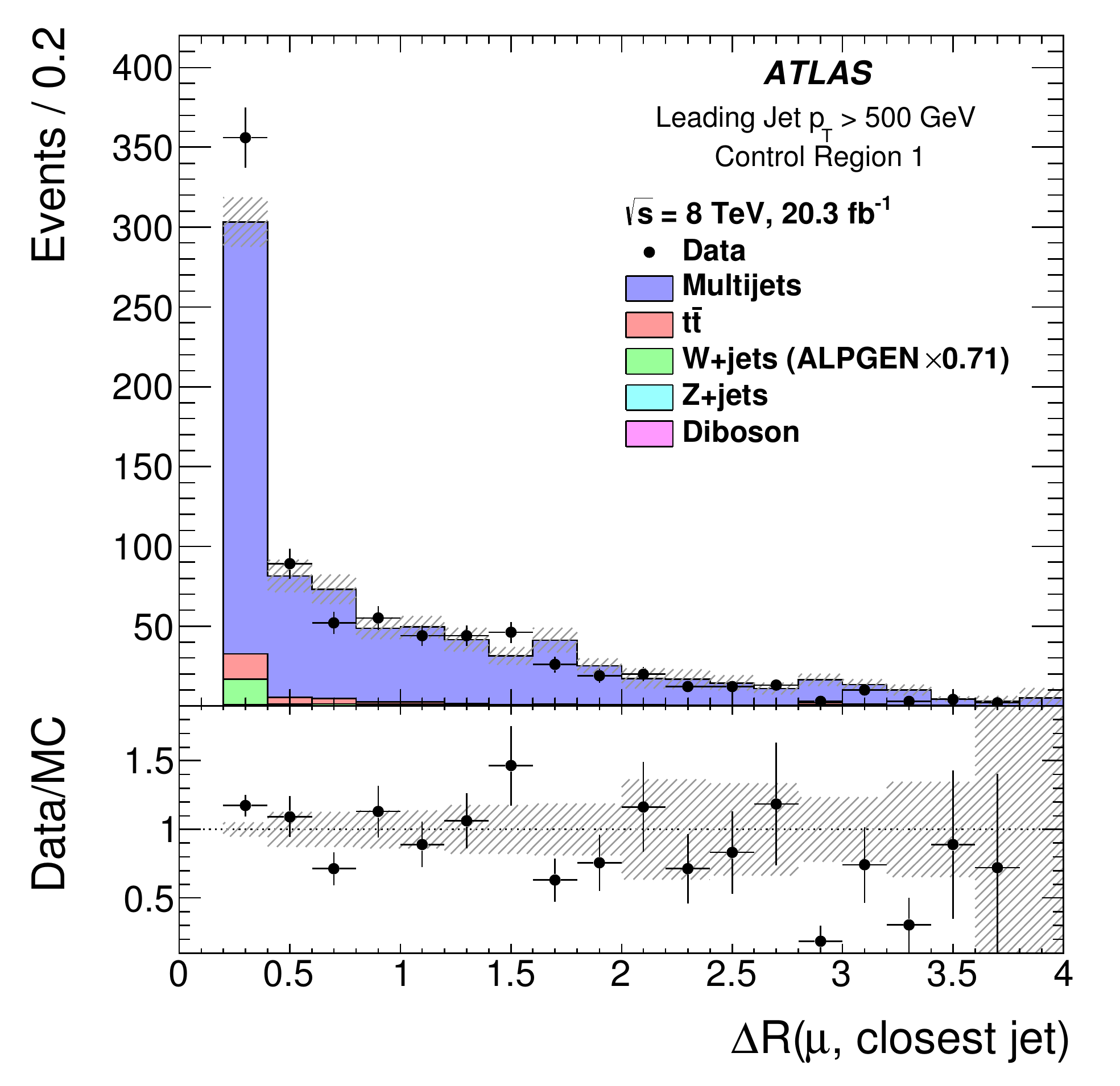}
  \includegraphics[width=3.1truein]{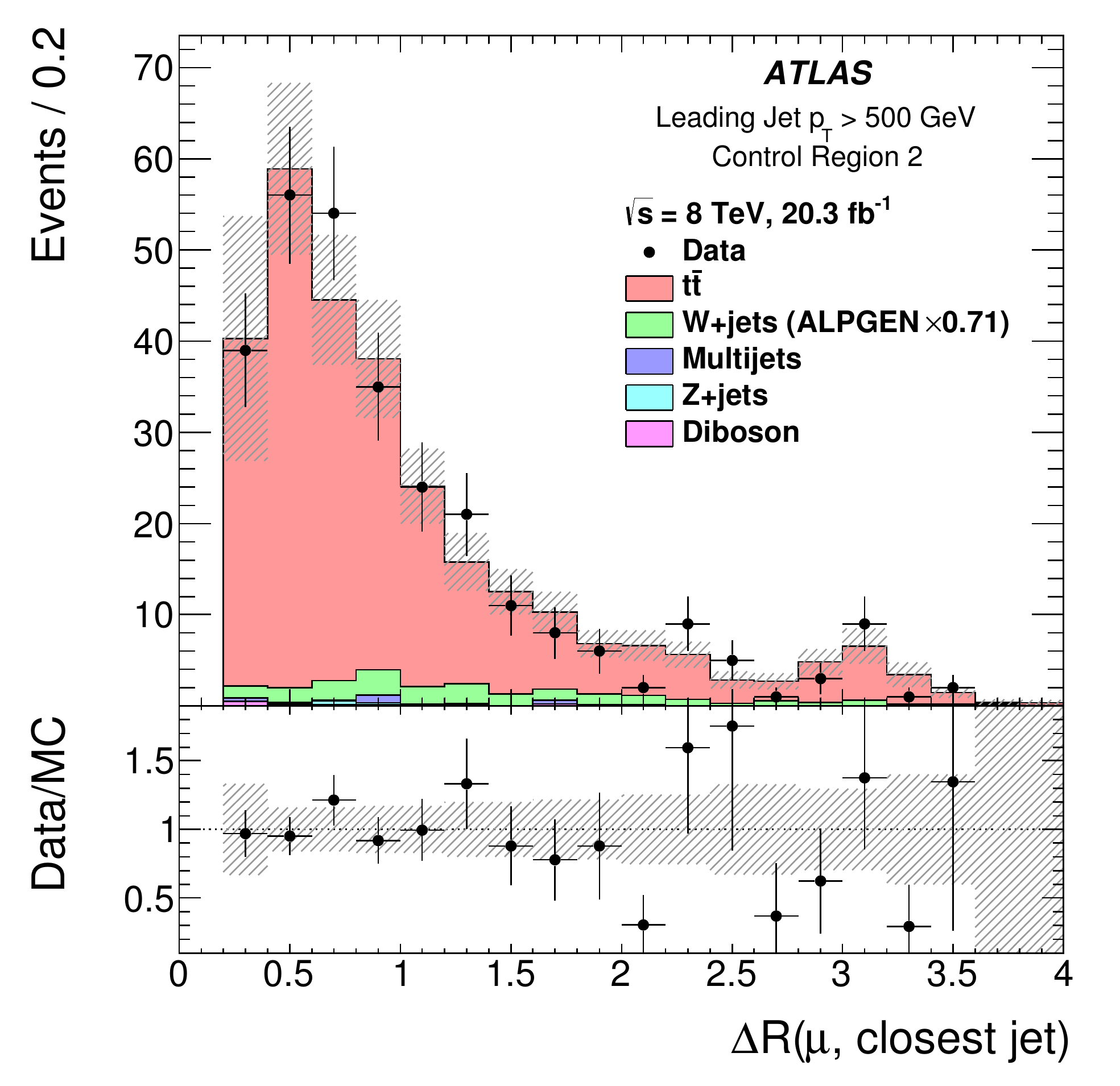} \\
  \includegraphics[width=3.1truein]{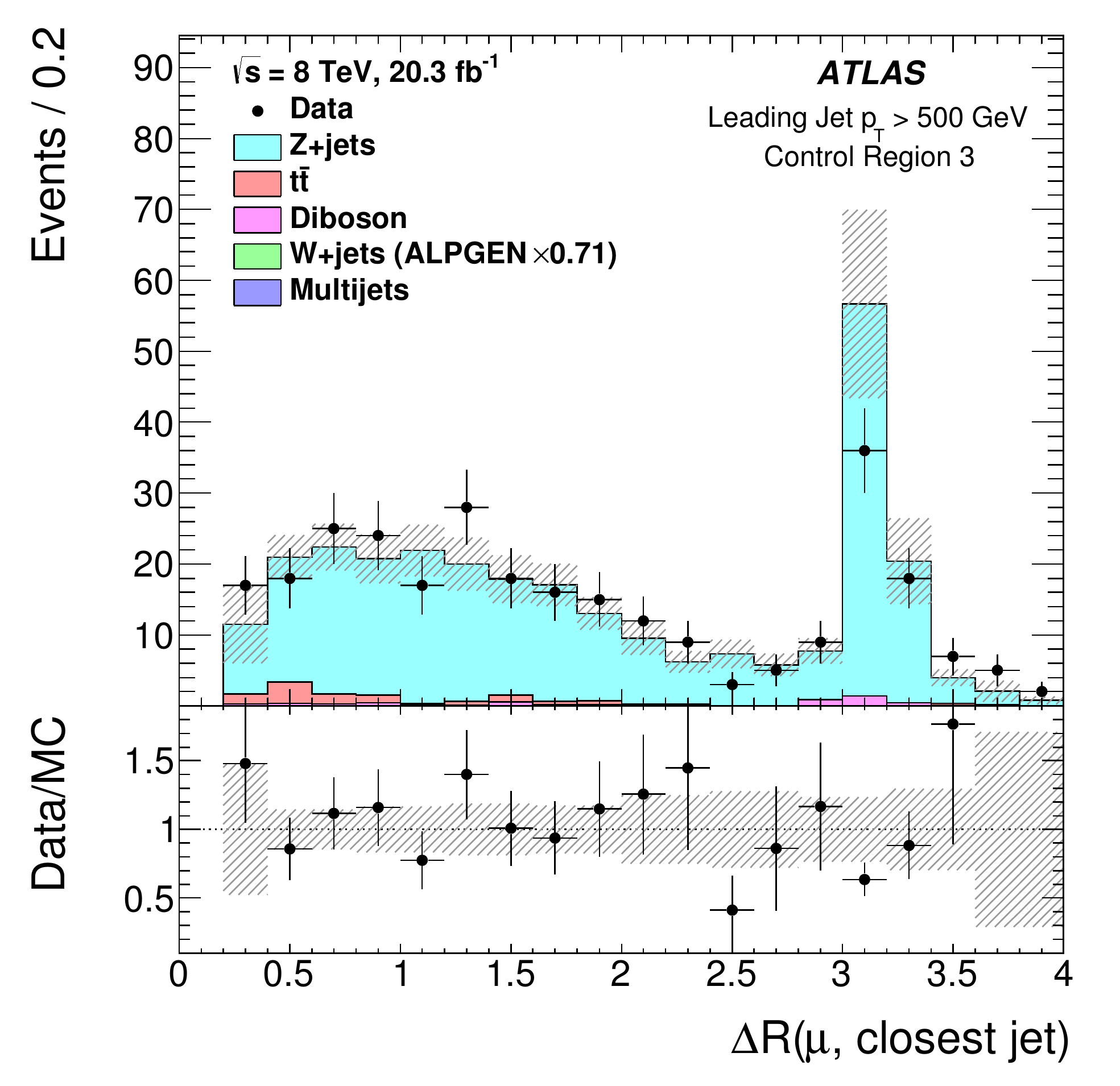}
  \caption{Comparisons between data and the predicted distribution from MC simulations of the angular separation between the muon and the closest jet in Control Region 1 (left), Control Region 2 (right) and Control Region 3 (bottom). The lower panels show the ratio of data to the predicted distribution. The error bars correspond to the statistical uncertainty and the shaded error band corresponds to the systematic uncertainties. The dijet, $\ttbar$ and \Zjets backgrounds have been scaled according to their respective control regions. The \Wjets signal has been scaled by $0.71$.
  \label{fig:event-selection:cr_deltar}}
\end{figure}

Using data from these control regions and the signal region, a scale factor is derived for each main background process and the \Wjets signal to correct the normalisation of the MC sample to that observed in data.
To ensure the scale factor is not affected by contamination from other backgrounds and the \Wjets signal, it is necessary to subtract the MC prediction for the contamination from the control region data.
As there is a circular dependency in using scaled MC predictions to derive new scalings, an iterative approach is applied.
First, the scale factors are derived with the contamination subtracted using the uncorrected normalisations.
Then the normalisations are updated with the scale factor corrections and the procedure to derive them is repeated.
Since the contamination in each of the regions is quite small, the scale factors converge very rapidly.
The dijet sample is scaled by $1.134 \pm 0.054$, the \ttbar sample is scaled by $0.861 \pm 0.061$, the \Zjets sample is scaled by $0.705 \pm 0.052$ and the \Wjets sample is scaled by $0.711 \pm 0.016$.
These uncertainties in the scale factors are due to the statistical uncertainty of the data and MC samples and are part of the overall uncertainties in the measurement detailed in \secref{systematics}.
However, the uncertainty in the \Wjets scale factor has no effect on the results of the measurement.
After the scale factors are applied, the MC predictions and observed distributions of the distance between the muon and the closest jet for each control region are shown in \figref{event-selection:cr_deltar}. The systematic uncertainties shown in \figref{event-selection:cr_deltar} correspond to those described in \secref{systematics}.

\section{Definition of observable and correction for detector effects}
\label{sec:unfolding}
The estimated background is subtracted from the data in the signal region and the resultant distribution of the distance $\Delta R$ between the muon and the closest jet is unfolded using an iterative Bayesian technique~\cite{2010arXiv1010.0632D} to correct for detector effects including both the efficiency of the selection criteria and the resolution of the angular separation between the muon and the nearest jet, where the former effect is dominant.
This technique is implemented within the RooUnfold framework~\cite{RooUnfold}.
A response matrix derived from MC simulation is used to correct the distribution from detector-level to particle-level.
The particle-level prediction from MC simulation is used as an initial prior during the first iteration of the unfolding.
Subsequent iterations use the previous iteration's unfolded distribution as a new prior.
A single iteration step is used, as this was found to be the optimal choice that minimised the combination of statistical fluctuation and the bias introduced by the prior of unfolded results.

The detector response and the combined efficiency of the trigger, reconstruction and the analysis selection for the \Wjets signal is obtained from MC simulation.
The fiducial selection applied to MC simulation is similar to the kinematic selection of the analysis.
Particle-level jets, built from stable final-state particles (defined as those with a proper lifetime $\tau$ corresponding to $c\tau \ge 10~\textnormal{mm}$~\cite{particleJets}) excluding muons and neutrinos, must satisfy $\pt > 100\GeV$ and $|\eta| < 2.1$.
Events are required to have at least one particle-level jet with $\pt > 500\GeV$ and a particle-level muon with a dressed\footnote{Photons that are contained in a cone of size $\Delta R = 0.1$ around the muon are summed and included as part of the muon energy.} $\pt > 25\GeV$ and $|\eta| < 2.4$.
No requirements on promptness are applied to the muons or the dressing photons.
Any additional muons that pass these requirements cause the event to be rejected.
Events where the distance between the muon and the closest jet $\DeltaR < 0.2$ are also rejected.
Unlike the analysis selection, there are no requirements on $b$-jets or electrons for the fiducial selection.

The unfolding to the fiducial region also corrects for events that do not pass the particle-level selection, but pass the detector-level selection.
Events in the fiducial signal region that arise from $W \rightarrow \tau \nu$ are also removed so that the cross-section is quoted exclusively for the muon decay channel.

\section{Systematic uncertainties}
\label{sec:systematics}
The dominant systematic uncertainties in the cross-section measurement arise from the uncertainties in the jet energy scale and the $b$-tagging efficiency.
For each systematic uncertainty, the selection criteria are re-applied, the control region normalisations are reassessed, and the unfolding procedure is repeated with the quantity under consideration varied by $\pm 1$ standard deviation.
The average of the up and down variations of the final cross-section measurement are summed in quadrature, as the variations are independent and not correlated.
This sum is then used as the full systematic uncertainty.
The systematic uncertainties in the measurement, grouped by source, are summarised in \tabref{systematics-xsecresult-sr2} for the inclusive cross-section, the collinear region ($0.2 < \DeltaR < 2.4$) and the back-to-back region ($\DeltaR > 2.4$).

Since the dijet, \ttbar and \Zjets simulated samples are scaled to data in their respective control regions, there is a systematic uncertainty in the scaling that arises from the statistical uncertainty in the data and the MC simulations in these control regions.
As the control region for dijets does not have the same kinematic selection as the signal region, there could be some bias due to mismodelling of the dijet kinematics in the simulated sample.
An uncertainty accounting for this is derived by varying the kinematic selection of the control region.

\begin{table}
  \begin{center}
    \begin{tabular}{|l|c|c|c|}
      \hline
      Systematic Source & {$0.2 < \DeltaR < 2.4$} & {$\DeltaR > 2.4$} & {Inclusive} \\
      \hline
      Scaling of dijets to data & 0.4\% & 0.1\% & 0.3\% \\
      Scaling of \ttbar to data & 0.6\% & 0.2\% & 0.5\% \\
      Scaling of \Zjets to data & 0.6\% & 0.3\% & 0.5\% \\
      Jet energy scale & 4.6\% & 5.8\% & 5.0\% \\
      $b$-tagging efficiency & 3.7\% & 1.2\% & 2.9\% \\
      Data/MC disagreement for dijets & 0.9\% & 0.6\% & 0.8\% \\
      Data/MC disagreement for \ttbar & 1.2\% & 0.4\% & 1.0\% \\
      Data/MC disagreement for \Zjets & 0.6\% & 1.5\% & 0.9\% \\
      Diboson background estimate & 2.2\% & 0.1\% & 1.5\% \\
      Unfolding dependence on prior & 1.1\% & 1.8\% & 1.3\% \\
      Muon momentum scale and resolution & 0.0\% & 0.1\% & 0.1\% \\
      Muon reconstruction efficiency & 0.4\% & 0.4\% & 0.4\% \\
      Muon trigger efficiency & 2.0\% & 1.9\% & 1.9\% \\
      Jet energy resolution & 0.6\% & 0.8\% & 0.6\% \\
      MC background statistical & 2.4\% & 1.8\% & 2.3\% \\
      MC response statistical & 1.7\% & 2.2\% & 1.9\% \\
      \hline
      Total systematic (excluding luminosity) & 7.6\% & 7.4\% & 7.3\% \\
      Luminosity & 1.9\% & 2.0\% & 2.0\% \\
      Data statistical & 2.7\% & 3.6\% & 2.2\% \\
      \hline
    \end{tabular}
    \caption{The systematic uncertainties in the cross-section measurement. Multiple independent components have been combined into groups of systematic uncertainties.
    \label{tab:systematics-xsecresult-sr2}}
  \end{center}
\end{table}

The uncertainty in the jet energy scale comprises 17 independent components~\cite{PERF-2012-01}.
Six of these are derived from various in situ analyses and two are related to the $\eta$ intercalibration of the jets.
There are also four components that account for the mismodelling of the $\pT$ response with respect to \pileup and three topology components that account for the dependence of the $\pT$-response uncertainty on the relative fractions of jets initiated by light quarks, gluons and $b$-quarks.

To correct the $b$-tagging efficiency in simulation to that observed in data, scale factors derived from in situ analyses are applied to the simulated samples~\cite{ATLAS-CONF-2014-046,ATLAS-CONF-2014-004}.
These have associated uncertainties.
The uncertainties for $b$-, $c$- and $\tau$-jets are assessed independently from those for light jets and the uncertainties in the efficiency scale factors are fully anti-correlated with those in the inefficiency scale factors.

In each control region, any disagreement between the $\DeltaR$ distributions for data and MC simulations is taken as a systematic uncertainty for the $\DeltaR$ prediction from that specific background in the signal region.
This introduces an additional data-driven systematic uncertainty to the dijet, \ttbar and \Zjets estimates for the $\DeltaR$ distribution.
Since the diboson background prediction is not constrained by data from a control region, an alternative prediction is obtained from a different simulated sample generated using \Sherpa.
The difference between these two predictions is taken as an uncertainty in the diboson background estimate.

The systematic uncertainty due to the dependence of the unfolding on the prior signal distribution, as obtained from MC simulations, is evaluated through a data-driven closure test.
The simulated signal sample is reweighted at particle-level such that the distribution of the fully simulated detector-level $\DeltaR$ more closely matches the observed data.
This reweighted simulated detector-level distribution is then unfolded and compared with the reweighted particle-level distribution.
Differences observed in this comparison are taken as a systematic uncertainty in the unfolding.
The uncertainty due to the dependence on the number of unfolding iteration steps was negligble.

Other smaller uncertainty contributions arise from the uncertainty in the integrated luminosity, the uncertainties in the muon momentum scale and resolution, muon reconstruction efficiency and trigger efficiency and the uncertainties in the jet energy resolution~\cite{Aad:2012ag}.
Uncertainties in the electron energy scale and resolution were evaluated but found to be negligible.

\section{Results}
\label{sec:results}
\begin{figure}
  \centering
  \includegraphics[width=4truein]{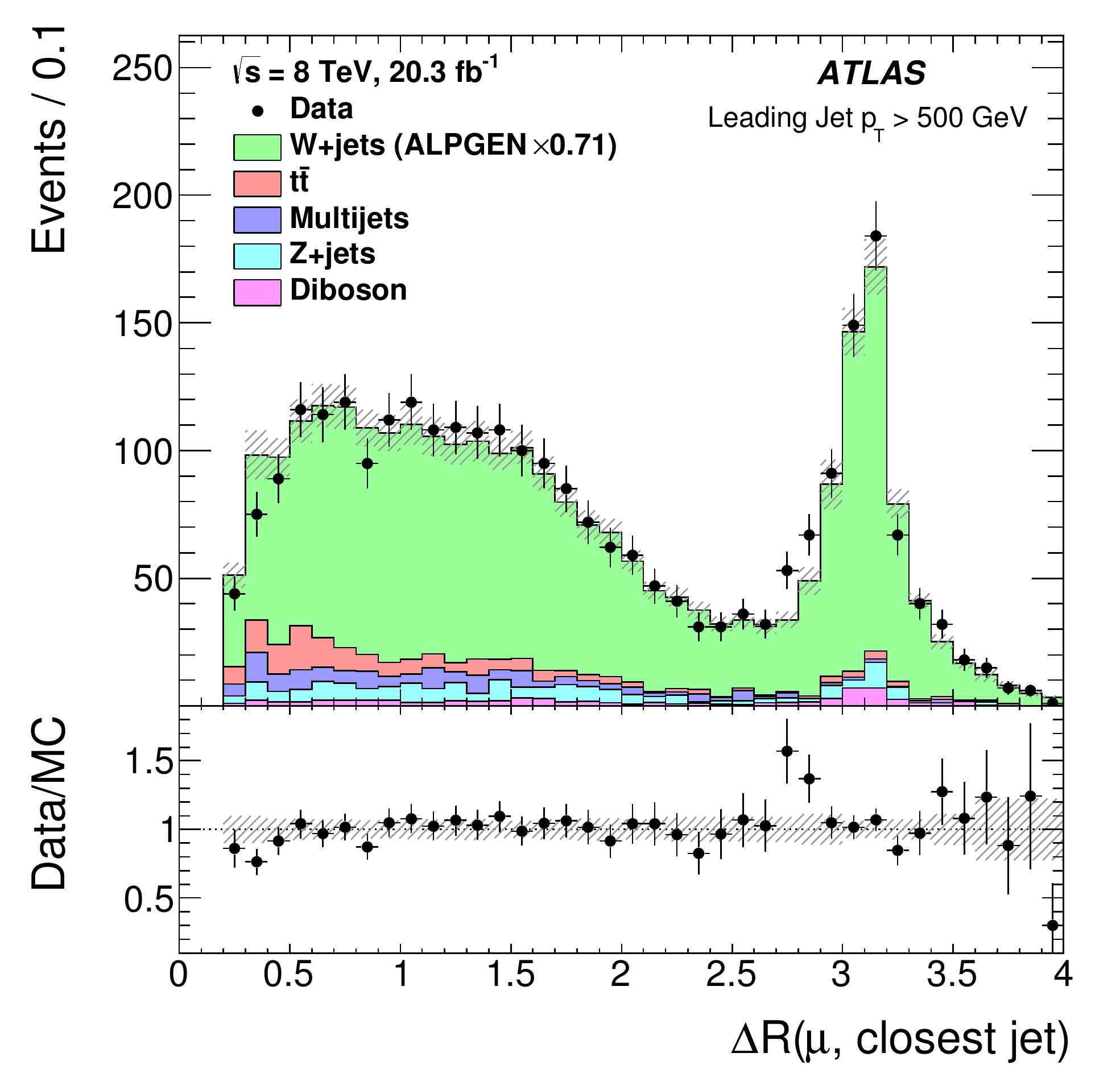}
  \caption{Predicted distribution from MC simulation of the angular separation between the muon and the closest jet and the observed distribution from data for the signal region. The lower panel shows the ratio of data to the predicted distribution. The error bars correspond to the statistical uncertainty and the shaded error band corresponds to the systematic uncertainties. The dijet, $\ttbar$ and \Zjets backgrounds have been scaled according to their respective control regions. The \Wjets signal has been scaled by $0.71$.
  \label{fig:sr2_deltar}}
\end{figure}

The number of events in the signal region observed in data is listed in \tabref{sr2-yields}, along with the composition of these events as predicted by MC simulation.
Numbers are given for the collinear region ($0.2 < \Delta R < 2.4$), the back-to-back region ($\Delta R > 2.4$), and the inclusive sample.
The uncorrected distributions of the reconstructed distance between the muon and the closest jet observed in data and predicted by MC simulations are shown in \figref{sr2_deltar} for the signal region.
In general the distributions agree within the uncertainties, except around $\Delta R = 2.8$ where there is a deficit and around the most collinear region of $\Delta R$ < 0.5 where there is a slight excess in the prediction from MC simulations.

\begin{table}
  \begin{center}
    \begin{tabular}{
    |l
    |r
    |r
    |r
    |
    }
      \hline
      Process & {$0.2 < \Delta R < 2.4$} & {$\Delta R > 2.4$} & {Inclusive} \\
      \hline
      Dijets & 5\% & 2\% & 4\% \\
      $\ttbar$ & 7\% & 2\% & 5\% \\
      \Zjets & 6\% & 4\% & 5\% \\
      Dibosons & 2\% & 4\% & 3\% \\
      \Wjets & 80\% & 88\% & 82\% \\
      \hline
      Data & 1907 & 833 & 2740 \\
      \hline
    \end{tabular}
    \caption{The number of events in the signal region observed in data, along with the composition of these events as predicted by MC simulation, split by the distance between the muon and the closest jet. The dijet, $\ttbar$ and \Zjets backgrounds have been scaled according to their respective control regions. The \Wjets signal has been scaled by $0.71$.
      \label{tab:sr2-yields}}
  \end{center}
\end{table}

\subsection{Differential cross-section measurement}
\label{sec:results:unfolded}

\begin{figure}
  \centering
  \includegraphics[width=4truein]{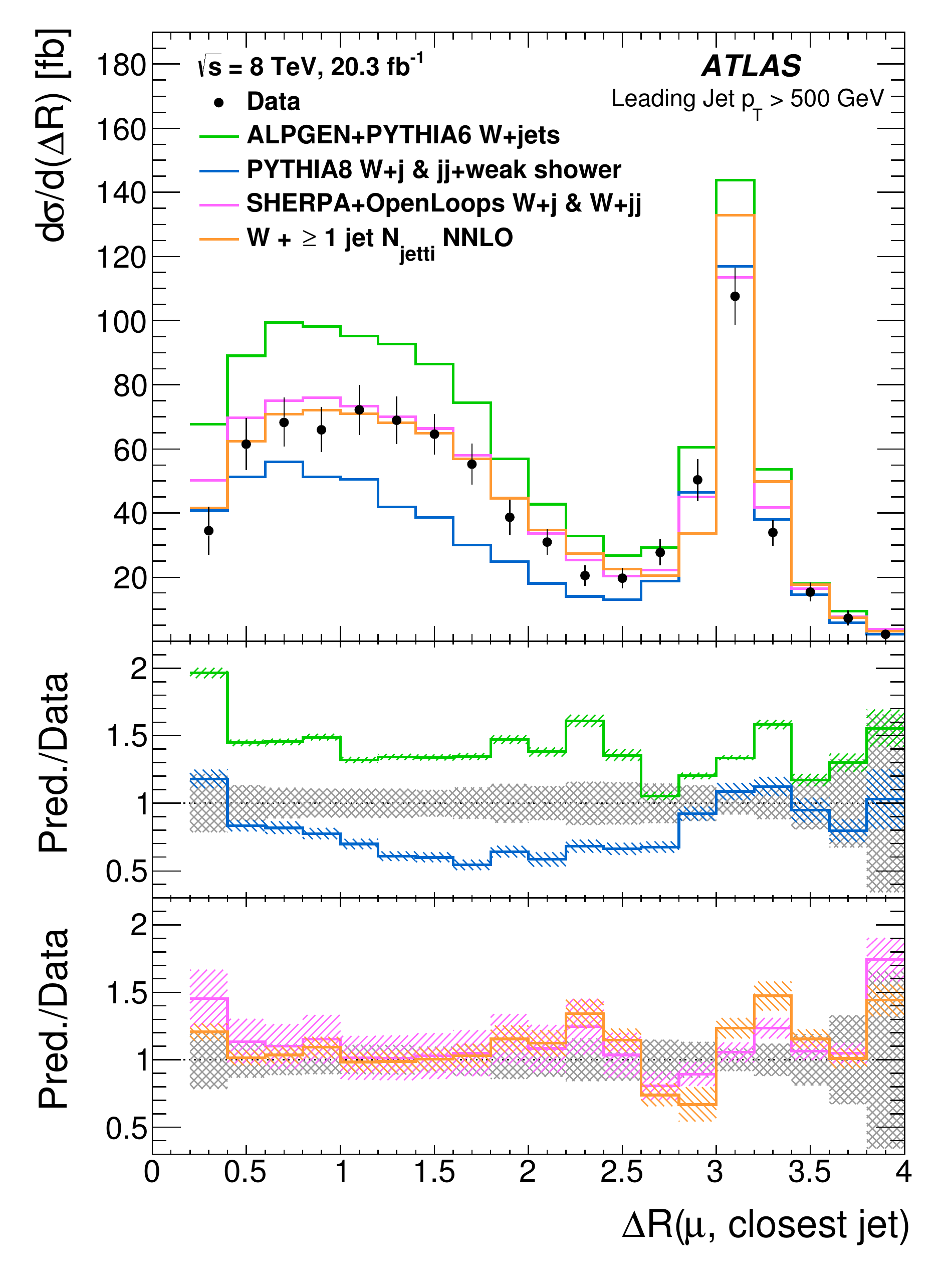}
  \caption{Unfolded distribution from background-subtracted data of the angular separation between the muon and the closest jet in the signal region along with several predictions from theory calculations. The lower panels show the ratio of the theory predictions to the unfolded data. The error bars in the upper panel and the grey shaded error bands in the lower ratio panels are the sum of the statistical and systematic uncertainties in the measurement. The shaded error band on the \AlpPythia calculation is statistical uncertainty, the band on the \PythiaEight calculation is statistical and PDF uncertainties and those on the SHERPA+OpenLoops and the $W~+ \ge 1~\textnormal{jet}$ $N_{\rm jetti}$ NNLO calculations are scale uncertainties.
  \label{fig:sr2_unfolded_deltar}}
\end{figure}

The differential cross-section of $W \rightarrow \mu \nu$ as a function of $\Delta R(\mu,\text{closest jet})$, obtained from the unfolded data of the signal region, is shown in \figref{sr2_unfolded_deltar}.
The measured total cross-sections for the inclusive case, in the collinear region and the back-to-back region are also listed in \tabrange{xsecresult}{xsecresult_backtoback}.

\begin{table}
  \begin{center}
    \begin{tabular}{
    |l
    | r @{ $\pm$ } l
    |
    }
      \hline
      Process & \multicolumn{2}{c|}{$\sigma\left(W(\rightarrow \mu\nu)~+ \ge 1~\textnormal{jet}\right)$ [fb]} \\
      \hline
      Data ($\sqrt{s} = 8$ TeV, $20.3~\textnormal{fb}^{-1}$) & 169.2 & 3.7 (stat.) $\pm$ 12.3 (syst.) $\pm$ 3.3 (lumi.) \\
      ALPGEN+PYTHIA6 $W$+jets & 236.6 & 1.1 (stat.) \\
      PYTHIA8 $W$+$j$ \& $jj$+weak shower & 134.8 & 0.9 (stat.) $\pm$ 7.3 (pdf) \\
      SHERPA+OpenLoops $W$+$j$ \& $W$+$jj$ & 183 & 25 (scale)\\
      $W~+ \ge 1~\textnormal{jet}$ $N_{\rm jetti}$ NNLO & 181 & 14 (scale)\\
      \hline
    \end{tabular}
    \caption{Cross-section for $W(\rightarrow \mu\nu)~+ \ge 1~\textnormal{jet}$ as measured in data and as predicted by various calculations.
      \label{tab:xsecresult}}
  \end{center}
\end{table}

\begin{table}
  \begin{center}
    \begin{tabular}{
    |l
    | r @{ $\pm$ } l
    |
    }
      \hline
      Process & \multicolumn{2}{c|}{$\sigma\left(W(\rightarrow \mu\nu)~+ \ge 1~\textnormal{jet},~0.2 < \Delta R < 2.4 \right)$ [fb]} \\
      \hline
      Data ($\sqrt{s} = 8$ TeV, $20.3~\textnormal{fb}^{-1}$) & 116.2 & 3.2 (stat.) $\pm$ 8.8 (syst.) $\pm$ 2.3 (lumi.) \\
      ALPGEN+PYTHIA6 $W$+jets & 167.1 & 0.9 (stat.) \\
      PYTHIA8 $W$+$j$ \& $jj$+weak shower & 83.4 & 0.7 (stat.) $\pm$ 4.4 (pdf) \\
      SHERPA+OpenLoops $W$+$j$ \& $W$+$jj$ & 128 & 20 (scale)\\
      $W~+ \ge 1~\textnormal{jet}$ $N_{\rm jetti}$ NNLO & 123 & 9 (scale)\\
      \hline
    \end{tabular}
    \caption{Cross-section for $W(\rightarrow \mu\nu)~+ \ge 1~\textnormal{jet}$ in the collinear ($0.2 < \Delta R < 2.4$) region as measured in data and as predicted by various calculations.
      \label{tab:xsecresult_collinear}}
  \end{center}
\end{table}

\begin{table}
  \begin{center}
    \begin{tabular}{
    |l
    | r @{ $\pm$ } l
    |
    }
      \hline
      Process & \multicolumn{2}{c|}{$\sigma\left(W(\rightarrow \mu\nu)~+ \ge 1~\textnormal{jet},~\Delta R > 2.4 \right)$ [fb]} \\
      \hline
      Data ($\sqrt{s} = 8$ TeV, $20.3~\textnormal{fb}^{-1}$) & 53.0 & 1.9 (stat.) $\pm$ 3.9 (syst.) $\pm$ 1.0 (lumi.) \\
      ALPGEN+PYTHIA6 $W$+jets & 69.5 & 0.6 (stat.) \\
      PYTHIA8 $W$+$j$ \& $jj$+weak shower & 51.4 & 0.6 (stat.) $\pm$ 2.9 (pdf) \\
      SHERPA+OpenLoops $W$+$j$ \& $W$+$jj$ & 55 & 5 (scale)\\
      $W~+ \ge 1~\textnormal{jet}$ $N_{\rm jetti}$ NNLO & 58 & 5 (scale)\\
      \hline
    \end{tabular}
    \caption{Cross-section for $W(\rightarrow \mu\nu)~+ \ge 1~\textnormal{jet}$ in the back-to-back ($\Delta R >2.4$) region as measured in data and as predicted by various calculations.
      \label{tab:xsecresult_backtoback}}
  \end{center}
\end{table}

The measurements are compared to several theory predictions.
The \AlpPythia \Wjets calculation and the normalisation $K$-factor used for this prediction are described in \secref{datamc} and the quoted uncertainties are the statistical uncertainties.
The $W$+$j$ and $jj$+weak shower calculation provided by \Pythia~v8.210, described in \secref{datamc}, is shown as well.
In this case, the \Wboson boson can either be produced by the matrix elements of the $\Wboson +  1$-jet final state or be emitted as electroweak final-state radiation in the parton shower of a dijet event.
The quoted uncertainties are the sums of the statistical uncertainties and the uncertainties from the CT10 NLO PDF set.
The data are compared to the nominal predictions from \AlpPythia and \PythiaEight.

The SHERPA+OpenLoops $W$+$j$ and $W$+$jj$ calculation incorporates NLO QCD and NLO EW corrections to both of these processes~\cite{2016JHEP...04..021K,2014arXiv1412.5157K,2002JHEP...02..044K,2008EPJC...53..501G,2012PhRvL.108k1601C,2014arXiv1407.0087D}.
In the high-\pt\ regime of the analysis, the NLO EW corrections can have significant effects -- up to 20\% -- across the $\Delta R$ distribution.
A second-jet veto is applied to the $W$+$j$ NLO predictions and this is then combined with the $W$+$jj$ NLO predictions.
The SHERPA+OpenLoops calculation also includes contributions from off-shell boson production and the sub-leading Born-level contributions ($O(\alpha^3)$ for $W$+$j$ and $O(\alpha_{\mathrm{S}} \alpha^3)$ for $W$+$jj$).
The NNPDF2.3QED NLO PDF~\cite{Ball:2013hta} is used.
Both the renormalisation and factorisation scales are set to $\mu_0 = 1/2 \left( \sqrt{m_{\mu v}^2 + (\pT^{\mu v})^2} + \Sigma_i \pT^{J_i} + \Sigma_i \pT^{\gamma_i} \right)$, where $m_{\mu v}$ and $\pT^{\mu v}$ are the mass and transverse momentum of the total four-momentum of the dressed muon and neutrino, $\pT^{J_i}$ is the transverse momentum of each jet, and $\pT^{\gamma_i}$ is the transverse momentum of each photon not used for dressing.
The quoted uncertainties are the scale uncertainties, where the renormalisation scale and the factorisation scale have been varied independently by a factor of two.

An NNLO QCD calculation, which includes up to $O(\alpha_{\mathrm{S}}^3)$, for the angular separation between the lepton from the \Wboson boson decay and the nearest jet in \Wjets events has recently become available~\cite{Boughezal:2015dva,Boughezal:2016yfp}.
This calculation, obtained from Ref.~\cite{Boughezal:2016dtm}, is denoted `$W~+ \ge 1~\textnormal{jet}$ $N_{\rm jetti}$ NNLO' here.
It uses a new technique based on $N$-jettiness~\cite{Gaunt:2015pea} to split the phase space for the real emission corrections.
It relies on the theoretical formalism provided in soft-collinear effective theory.
The calculation uses the CT14 NNLO PDF~\cite{Dulat:2015mca} and $\mu_0 = \sqrt{m_{\ell v}^2 + \Sigma_i (\pT^{J_i})^2}$, where $m_{\ell v}$ is the invariant mass of the lepton and neutrino and $\pT^{J_i}$ is the transverse momentum of each jet, is used for both the renormalisation and factorisation scale.
The quoted uncertainties are the scale uncertainties, where the renormalisation scale and the factorisation scale have been varied independently by a factor of two.
The resulting partonic final state is clustered using the \antikt\ jet algorithm with $R = 0.4$.
No non-perturbative corrections are applied.
The selections used in the calculation,
\begin{equation}
  \begin{split}
    \pT^{\text{jet}} &> 100 \GeV, \;\;\; |\eta^{\text{jet}}|<2.1,\;\;\; \pT^{\text{leading jet}}>500 \GeV, \\
    \pT^{\ell} &> 25 \GeV, \;\;\; |\eta^{\ell}| < 2.5,
  \end{split}
\label{eq:cuts2}
\end{equation}
are the same as the ones used for the measurement except for the muon pseudorapidity ($|\eta| < 2.5$ instead of $|\eta| < 2.4$).
The effect of this difference in muon pseudorapidity is evaluated using the \AlpPythia \Wjets sample and a correction factor accounting for this, which is less than 4\% across the entire distribution, is applied.
The calculated cross-sections obtained at LO, NLO and NNLO without the muon pseudorapidity correction are shown in \tabref{fidboost}.
The scale uncertainty decreases from $\sim\pm 20\%$ at NLO to $+3\%/{-7}\%$ at NNLO.

{\renewcommand{\arraystretch}{1.2}
\begin{table}[h]
\begin{center}
\begin{tabular}{|c|c|c|c|c|c|}
\hline
 & $\sigma_{\LO}$ [fb] & $\sigma_{\NLO}$ [fb] & $\sigma_{\NNLO}$ [fb] \\
 \hline
8 TeV & $57^{+13}_{-10}$ & $160^{+35}_{-27}$ & $187^{+5}_{-12}$ \\
\hline
\end{tabular}
\caption{Fiducial \Wjets cross-sections for the selection criteria of \equref{cuts2} at LO, NLO and NNLO in QCD from Ref.~\cite{Boughezal:2016dtm}. The uncertainties shown are the scale uncertainties.}
\label{tab:fidboost}
\end{center}
\end{table}
}

The comparison of the data to \AlpPythia in \figref{sr2_unfolded_deltar} shows good shape agreement to within uncertainties, except at very low $\DeltaR$, but \AlpPythia predicts a significantly higher integrated cross-section.
The comparison to \PythiaEight at high $\DeltaR$, where it is dominated by back-to-back \Wjets production in which the \Wboson boson is balanced by the hadronic recoil system, shows much better agreement.
At smaller $\DeltaR$, where the collinear process dominates, neither the shape nor the overall cross-section agree.
The comparisons to SHERPA+OpenLoops and $W~+ \ge 1~\textnormal{jet}$ $N_{\rm jetti}$ NNLO show much better agreement across the entire distribution.

\subsection{Enhancement of the collinear fraction with jet $\pT$}
\label{sec:results:enhancementpt}

\begin{figure}
  \centering
  \includegraphics[width=4truein]{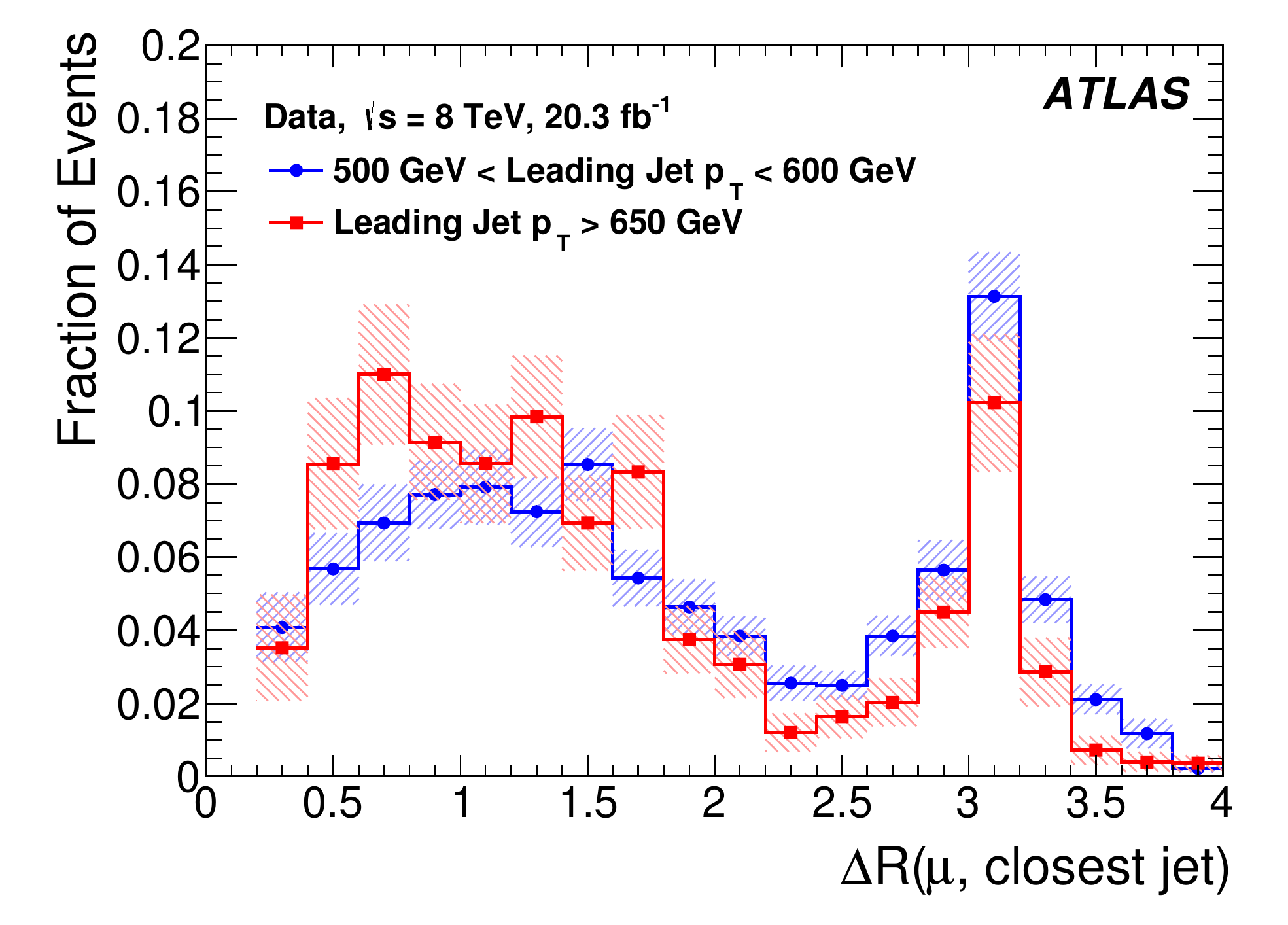}
  \caption{Unfolded distribution from background-subtracted data of the angular separation between the muon and the closest jet for events with $500\GeV < \pT^{\textnormal{leading jet}} < 600\GeV$ (blue circles) and $\pT^{\textnormal{leading jet}} > 650\GeV$ (red squares) from the signal region. Distributions are normalised to unity. The shaded error band on the unfolded measurement corresponds to the sum of the statistical and systematic uncertainties.
  \label{fig:sr2ab_unfolded_deltar_normalized}}
\end{figure}

\begin{figure}
  \centering
  \subfigure[]{
    \includegraphics[width=0.45\textwidth]{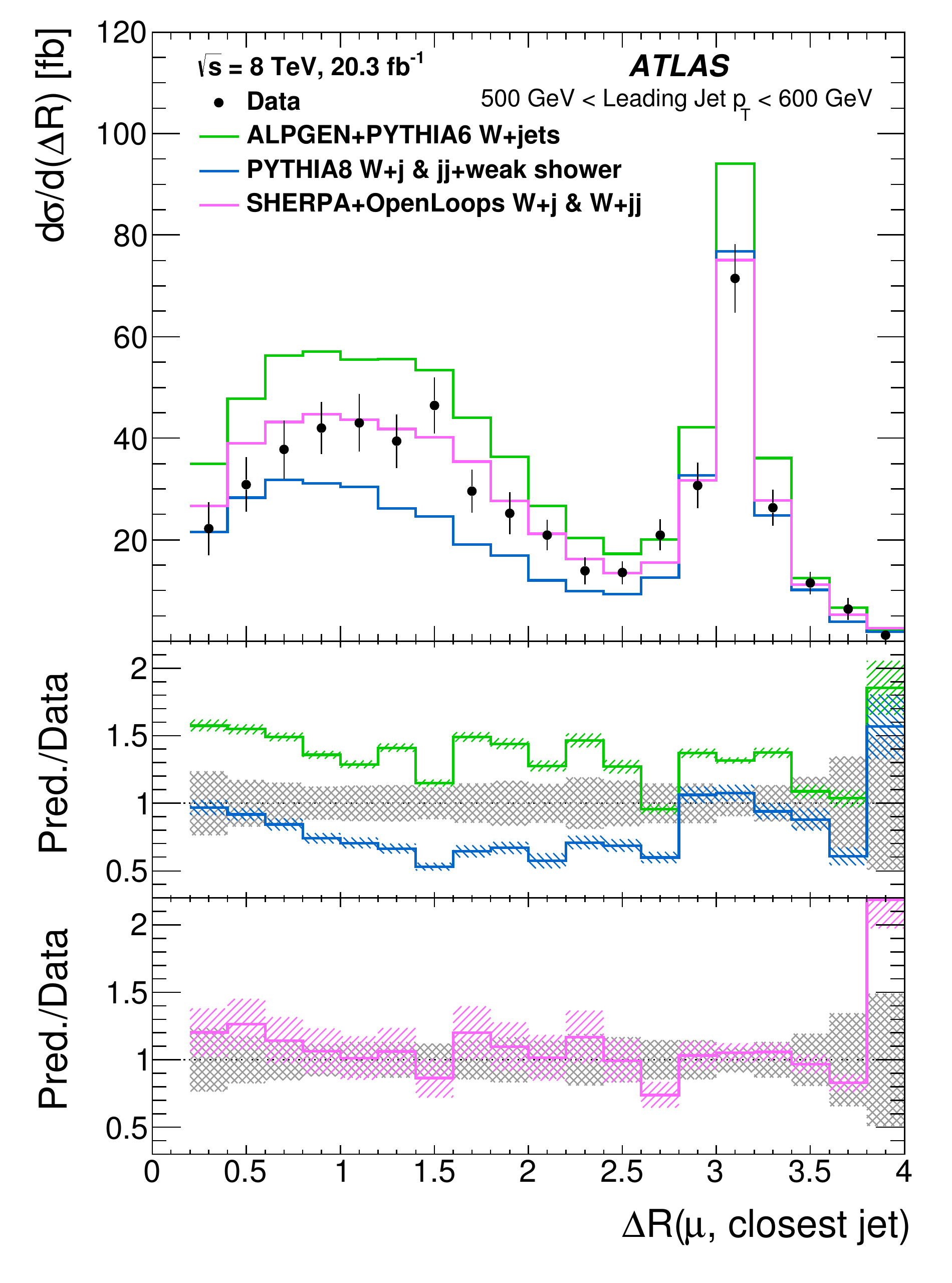}
  }
  \subfigure[]{
    \includegraphics[width=0.45\textwidth]{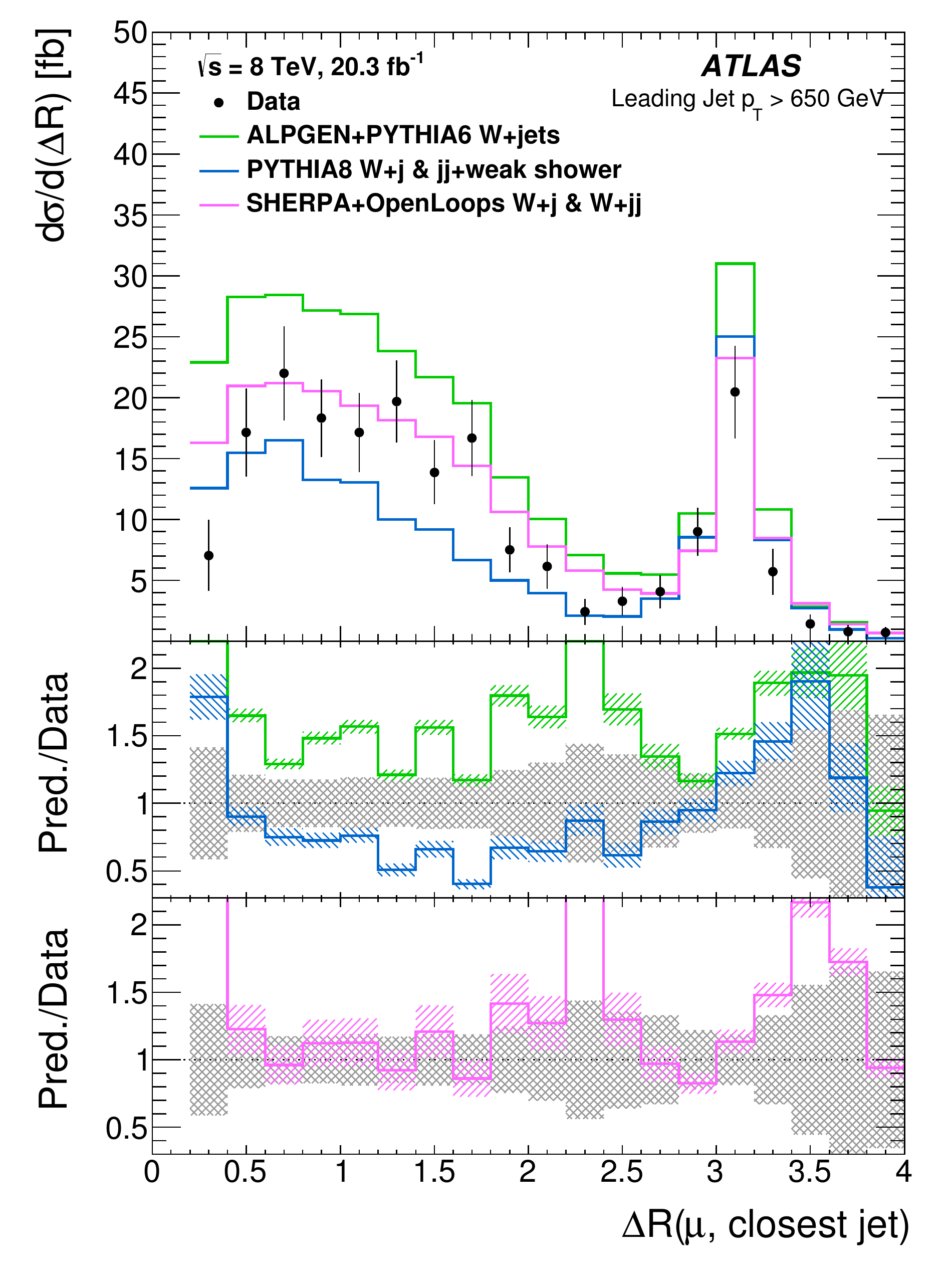}
  }
  \caption{Unfolded distribution from background-subtracted data of the angular separation between the muon and the closest jet in the signal region along with several predictions from theory calculations for events with (a) $500\GeV < \pT^{\textnormal{leading jet}} < 600\GeV$ and (b) $\pT^{\textnormal{leading jet}} > 650\GeV$. The lower panels show the ratio of the theory predictions to the unfolded data. The error bars in the upper panel and the grey shaded error bands in the lower ratio panels are the sum of the statistical and systematic uncertainties in the measurement. The shaded error band on the \AlpPythia calculation is statistical uncertainty, the band on the \PythiaEight calculation is statistical and PDF uncertainties and the band on the SHERPA+OpenLoops is scale uncertainty.
  \label{fig:sr2ab_unfolded_deltar}}
\end{figure}

The events in the signal region are further divided into two categories based on the transverse momentum of the leading jet: $500\GeV < \pT^{\textnormal{leading jet}} < 600\GeV$ and $\pT^{\textnormal{leading jet}} > 650\GeV$.
For each of these two categories, the data distribution is unfolded.
The $50\GeV$ gap between the two categories reduces the migration of events from one category to the other during unfolding.
The resulting normalised differential \Wjets cross-section is shown in \figref{sr2ab_unfolded_deltar_normalized}.
As the leading-jet $\pT$ increases, the fraction of events in the lower $\DeltaR$ (collinear) region increases and the fraction in the higher $\DeltaR$ (back-to-back \Wjets) region decreases.
This may be interpreted as an increase in the collinear \Wboson emission probability  as the jets become more energetic.
With higher $\pT$ the collinear peak is shifted to smaller $\Delta R$.
This is also understood since the mass of the \Wboson boson becomes proportionally smaller compared to the energy of the jet.
The full measurement results are shown in \figref{sr2ab_unfolded_deltar}.
The comparison to theory predictions shows results similar to the ones obtained for $\pT^{\textnormal{leading jet}} > 500\GeV$ in \secref{results:unfolded}.

\FloatBarrier
\section{Conclusions}
\label{sec:conclusions}
The cross-section for $W \rightarrow \mu\nu$ in association with at least one very high transverse momentum jet is measured as a function of the angular distance between the muon from the \Wboson boson decay and the closest jet.
This measurement utilises data recorded by the ATLAS detector from $pp$ collisions at $\sqrt{s}=8$~TeV at the LHC, corresponding to \intluminoerr of integrated luminosity.
These results are relevant to understanding the contribution of real \Wboson emissions from high-\pt light partons to \Wjets processes.

Comparisons to a variety of MC generators and theoretical calculations show varying levels of agreement.
\AlpPythia overestimates the total cross-section, whereas \PythiaEight, which is modified to explicitly include the process of \Wboson boson emission, disagrees with the measurement in the collinear region ($\Delta R < 2.4$).
On the other hand, agreement with the SHERPA+OpenLoops NLO QCD+EW calculation and the $W~+ \ge 1~\textnormal{jet}$ $N_{\rm jetti}$ NNLO calculation in Ref.~\cite{Boughezal:2016dtm} is well within the systematic and statistical uncertainties of the predictions and the measurement.

This measurement has implications for Monte Carlo programs that incorporate real \Wboson boson emission, a process which is only just now being probed directly at the energy of the LHC. The rate of this process increases with jet \pt\ and thus also with centre-of-mass energy, and will therefore play a significant role in \Wjets measurements at high \pt, vector-boson scattering measurements, and even QCD multijet measurements at very large dijet invariant masses where the corrections due to real boson emission are significant.

Lastly, the potential is high for this process to mimic the signatures of a highly Lorentz-boosted top quark. The importance of such signatures in the search for new physics at the LHC necessitates a thorough understanding of processes such as the one measured in detail in this paper. As the physics programmes of the LHC experiments extend into new territories in terms of both the centre-of-mass energy and integrated luminosity, these once rare processes will become a ubiquitous consideration.

\section{Acknowledgements}
\label{sec:ack}
% Acknowledgements for papers with collision data
% Version 2-Aug-2016

% Standard acknowledgements start here
%----------------------------------------------
We thank CERN for the very successful operation of the LHC, as well as the
support staff from our institutions without whom ATLAS could not be
operated efficiently.

We acknowledge the support of ANPCyT, Argentina; YerPhI, Armenia; ARC, Australia; BMWFW and FWF, Austria; ANAS, Azerbaijan; SSTC, Belarus; CNPq and FAPESP, Brazil; NSERC, NRC and CFI, Canada; CERN; CONICYT, Chile; CAS, MOST and NSFC, China; COLCIENCIAS, Colombia; MSMT CR, MPO CR and VSC CR, Czech Republic; DNRF and DNSRC, Denmark; IN2P3-CNRS, CEA-DSM/IRFU, France; GNSF, Georgia; BMBF, HGF, and MPG, Germany; GSRT, Greece; RGC, Hong Kong SAR, China; ISF, I-CORE and Benoziyo Center, Israel; INFN, Italy; MEXT and JSPS, Japan; CNRST, Morocco; FOM and NWO, Netherlands; RCN, Norway; MNiSW and NCN, Poland; FCT, Portugal; MNE/IFA, Romania; MES of Russia and NRC KI, Russian Federation; JINR; MESTD, Serbia; MSSR, Slovakia; ARRS and MIZ\v{S}, Slovenia; DST/NRF, South Africa; MINECO, Spain; SRC and Wallenberg Foundation, Sweden; SERI, SNSF and Cantons of Bern and Geneva, Switzerland; MOST, Taiwan; TAEK, Turkey; STFC, United Kingdom; DOE and NSF, United States of America. In addition, individual groups and members have received support from BCKDF, the Canada Council, CANARIE, CRC, Compute Canada, FQRNT, and the Ontario Innovation Trust, Canada; EPLANET, ERC, FP7, Horizon 2020 and Marie Sk{\l}odowska-Curie Actions, European Union; Investissements d'Avenir Labex and Idex, ANR, R{\'e}gion Auvergne and Fondation Partager le Savoir, France; DFG and AvH Foundation, Germany; Herakleitos, Thales and Aristeia programmes co-financed by EU-ESF and the Greek NSRF; BSF, GIF and Minerva, Israel; BRF, Norway; Generalitat de Catalunya, Generalitat Valenciana, Spain; the Royal Society and Leverhulme Trust, United Kingdom.

The crucial computing support from all WLCG partners is acknowledged gratefully, in particular from CERN, the ATLAS Tier-1 facilities at TRIUMF (Canada), NDGF (Denmark, Norway, Sweden), CC-IN2P3 (France), KIT/GridKA (Germany), INFN-CNAF (Italy), NL-T1 (Netherlands), PIC (Spain), ASGC (Taiwan), RAL (UK) and BNL (USA), the Tier-2 facilities worldwide and large non-WLCG resource providers. Major contributors of computing resources are listed in Ref.~\cite{ATL-GEN-PUB-2016-002}.
%----------------------------------------------

\clearpage

\bibliographystyle{bibtex/bst/atlasBibStyleWithTitle}
\bibliography{../COMMON/collinearW.bib,bibtex/bib/ATLAS,bibtex/bib/ConfNotes,bibtex/bib/PubNotes}

\newpage % ATLAS Collaboration author list
% Data extracted on 12-Aug-2016 for paper reference STDM-2015-16
%\documentclass[11pt]{article}
%\usepackage{a4wide}\begin{document}
\begin{flushleft}
{\Large The ATLAS Collaboration}

\bigskip

M.~Aaboud$^\textrm{\scriptsize 136d}$,
G.~Aad$^\textrm{\scriptsize 87}$,
B.~Abbott$^\textrm{\scriptsize 114}$,
J.~Abdallah$^\textrm{\scriptsize 8}$,
O.~Abdinov$^\textrm{\scriptsize 12}$,
B.~Abeloos$^\textrm{\scriptsize 118}$,
R.~Aben$^\textrm{\scriptsize 108}$,
O.S.~AbouZeid$^\textrm{\scriptsize 138}$,
N.L.~Abraham$^\textrm{\scriptsize 152}$,
H.~Abramowicz$^\textrm{\scriptsize 156}$
% ...
\bigskip
\\
% ...
$^{8}$ Department of Physics, The University of Texas at Arlington, Arlington TX, United States of America\\
% ...
$^{12}$ Institute of Physics, Azerbaijan Academy of Sciences, Baku, Azerbaijan\\
% ...
$^{87}$ CPPM, Aix-Marseille Universit{\'e} and CNRS/IN2P3, Marseille, France\\
% ...
$^{108}$ Nikhef National Institute for Subatomic Physics and University of Amsterdam, Amsterdam, Netherlands\\
% ...
$^{114}$ Homer L. Dodge Department of Physics and Astronomy, University of Oklahoma, Norman OK, United States of America\\
% ...
$^{118}$ LAL, Univ. Paris-Sud, CNRS/IN2P3, Universit{\'e} Paris-Saclay, Orsay, France\\
% ...
$^{136}$ $^{(a)}$ Facult{\'e} des Sciences Ain Chock, R{\'e}seau Universitaire de Physique des Hautes Energies - Universit{\'e} Hassan II, Casablanca; $^{(b)}$ Centre National de l'Energie des Sciences Techniques Nucleaires, Rabat; $^{(c)}$ Facult{\'e} des Sciences Semlalia, Universit{\'e} Cadi Ayyad, LPHEA-Marrakech; $^{(d)}$ Facult{\'e} des Sciences, Universit{\'e} Mohamed Premier and LPTPM, Oujda; $^{(e)}$ Facult{\'e} des sciences, Universit{\'e} Mohammed V, Rabat, Morocco\\
% ...
$^{138}$ Santa Cruz Institute for Particle Physics, University of California Santa Cruz, Santa Cruz CA, United States of America\\
% ...
$^{152}$ Department of Physics and Astronomy, University of Sussex, Brighton, United Kingdom\\
% ...
$^{156}$ Raymond and Beverly Sackler School of Physics and Astronomy, Tel Aviv University, Tel Aviv, Israel\\
% ...
$^{a}$ Also at Department of Physics, King's College London, London, United Kingdom\\
$^{b}$ Also at Institute of Physics, Azerbaijan Academy of Sciences, Baku, Azerbaijan\\
$^{c}$ Also at Novosibirsk State University, Novosibirsk, Russia\\
$^{d}$ Also at TRIUMF, Vancouver BC, Canada\\
$^{e}$ Also at Department of Physics {\&} Astronomy, University of Louisville, Louisville, KY, United States of America\\
% ...
\end{flushleft}

%\end{document}
% Created with xml2latex.py

\end{document}